# Wiener Reconstruction, SVD and 'Optimal' Functional Bases: Application For Redshift Galaxy Catalogs

Saleem Zaroubi
*Astronomy Department and Center for Particle Astrophysics*
*University of California, Berkeley, U.S.A.*

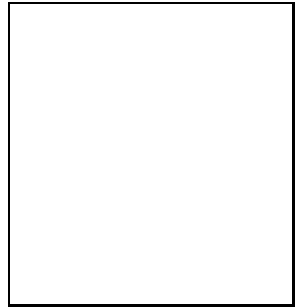

**Abstract**

The formalism of Wiener filtering is applied to reconstruct, in terms of spherical harmonics, the projected $4\pi$ galaxy distribution of a mock IRAS 1.2 Jy catalog with a 'Zone of Avoidance' $|b| = 15°$. The Singular Value Decomposition algorithm is utilized as an extra-regularizer in order to stabilize the inversion. This case sheds light on the amount of useful information contained in the data, and also suggests a method for obtaining an 'optimal' functional basis, based on our *prior* knowledge, for the purposes of likelihood analysis and Wiener reconstruction of cosmological data sets. The applicability of such a functional basis for redshift galaxy catalogs is also discussed

## 1 Introduction

Galaxy catalogs provide the main data base for the purpose of mapping the (continuous) cosmological dynamical fields (density, velocity and gravitational potential), which in turn offer a probe of the early universe and the nature of the primordial perturbation field. However, continuous density maps, constructed from the discrete galaxy distribution, suffer from Poisson 'shot-noise'. This problem become much more severe in the case of magnitude limited samples, where the selection function magnifies the shot-noise contribution as the probability of observing galaxies decreases.
Two other problems appear in analyzing the distribution of galaxies. First, the incomplete sky coverage, due to the obscuration by gas and dust in our local galaxy, the so-called 'Zone of Avoidance' (ZOA), poses an obstacle in mapping the whole-sky distribution. Second, the peculiar velocities in redshift surveys give a distorted picture of the galaxy distribution, even in the linear regime of gravitational instability.
Recovering the underlying field from the measured field can be viewed as an inversion problem in the following sense. Consider the case in which the observed data points, $\mathbf{d} = \{d_i\}(i = 1, \ldots, M)$ are given by a linear convolution or mapping of the underlying field, $\mathbf{s} = \{s_\alpha\}(\alpha = 1, \ldots, N)$, namely,

$$\mathbf{d} = \mathbf{R}\,\mathbf{s} + \boldsymbol{\epsilon}, \tag{1}$$

where $\mathbf{R}$ is an $M \times N$ matrix which represents a response function and $\boldsymbol{\epsilon} = \{\epsilon_i\}$ $(i = 1, \ldots, M)$, is the statistical uncertainty vector associated with the data. Here the standard notion of the response function is extended to include any theoretical relationship between two fields (*e.g.*, the matrix which relates the redshift space density to the real space density [4] and [19]).

Mathematically, the act of reconstruction amounts to solving equation 1 for $\mathbf{s}$ given $\mathbf{d}$ where $\mathbf{R}$ is assumed to be known. However, direct inversion of equation 1, *i.e.*, $\mathbf{s} = \mathbf{R}^{-1}\mathbf{d}$, suffers from several problems, the most important of which are: First, the number of independent data points is often smaller than the number of degrees of freedom (*d.o.f.*) of the underlying field, and therefore the data do not contain enough information to constrain all of the *d.o.f.*. Second, the inversion might be unstable due to the existence of statistical noise.

As a result of these potential pitfalls of direct inversion, the technique of Wiener filtering (*c.f.*, [14], [20]) as a regularization method is presented and applied for the case of the projected mock IRAS 1.2 Jy catalog with $|b| = 15°$ ZOA.

The Singular Value Decomposition algorithm (SVD) [13] as an extra-regularizer is also applied. An extension of the SVD algorithm is presented and discussed in the context of Wiener reconstruction and maximum likelihood analysis of 3-D redshift catalogs ([20]).

## 2  Wiener Filter

The derivation of the Wiener filter (WF) requires *prior* knowledge of the the first two moments of the underlying field, namely, $\langle \mathbf{s} \rangle$ (taken in what follows to be 0 for simplicity), and its covariance matrix, $\mathbf{S} = \langle \mathbf{s}\,\mathbf{s}^\dagger \rangle \equiv \{\langle s_i\, s_j^* \rangle\}$. (see [9] [4] and [3] for various applications of WF). Define an estimator of the underlying field, $\mathbf{s}^{\mathrm{MV}}$ (MV stands for minimal variance), as the linear combination of the data, $\mathbf{s}^{\mathrm{MV}} = \mathbf{F}\,\mathbf{d}$, where $\mathbf{F}$ is an $N \times M$ matrix,, chosen to minimize the variance of the residual $\mathbf{r}$ defined by

$$\langle \mathbf{r}\,\mathbf{r}^\dagger \rangle = \langle (\mathbf{s} - \mathbf{s}^{\mathrm{MV}})(\mathbf{s}^\dagger - \mathbf{s}^{\mathrm{MV}\dagger}) \rangle. \tag{2}$$

Carrying out the minimization of equation 2 with respect to $\mathbf{F}$ one finds the so-called WF,

$$\mathbf{F} = \langle \mathbf{s}\,\mathbf{d}^\dagger \rangle \langle \mathbf{d}\,\mathbf{d}^\dagger \rangle^{-1} = \mathbf{S}\mathbf{R}^\dagger(\mathbf{R}\,\mathbf{S}\,\mathbf{R}^\dagger + \mathbf{N})^{-1}, \tag{3}$$

where $\mathbf{N} \equiv \langle \boldsymbol{\epsilon}\,\boldsymbol{\epsilon}^\dagger \rangle$ is the noise correlation matrix, which is assumed to be uncorrelated with the underlying field. WF includes two operations: *inversion* of the response function and *suppression* of the shot noise roughly by the ratio of $\frac{prior}{prior + noise}$. Note that this ratio is less than unity, and therefore the method can not be used iteratively as successive applications of the WF would drive the recovered field to zero (see [20] for more details).

While earlier applications of the WF have focused on estimation, namely suppressing the noise in the measured quantities, one can extend the application of the WF to predict the values of unmeasured quantities, such as the density field in un-sampled regions of space, or to deconvolve blurred data. Within the context of linear gravitational instability theory, the WF can also be used to perform dynamical reconstruction of one field which is dynamically related to some other observed field. This is the case, for example, of the reconstruction of the real space galaxy distribution from its redshift distribution.

## 2.1 Conditional Probability

The standard model of cosmology assumes that the primordial perturbation field is Gaussian, and therefore on large scales where the fluctuations are still small the perturbations field will be very close to Gaussian, even in the present epoch. The statistical properties of any Gaussian random field depend only on its two-point covariance matrix $\mathbf{S}$ [1]. If the noise is also a Gaussian field, then, since it is independent, one can construct simple estimates of the underlying field $\mathbf{s}$, based on the knowledge of the full probability distribution function.

In our case, if both the signal and the noise are Gaussian, the conditional mean value of the field given the data can also serve as an estimator of $\mathbf{s}$, $\int \mathbf{s} P(\mathbf{s}|\mathbf{d})\, d\mathbf{s}$, which yields $\mathbf{s}^{mean} = \mathbf{s}^{MV}$. $\mathbf{s}^{MV}$ coincides also with the maximum *a posteriori* estimate (MAP) of the field; namely the one that maximizes the conditional probability $P(\mathbf{s}|\mathbf{d})$. Therefore for Gaussian fields WF coincides with the mean and the MAP estimators, *i.e.*, $\mathbf{s}^{MV} = \mathbf{s}^{mean} = \mathbf{s}^{MAP}$. The residual from the mean coincides with $\mathbf{r}$, and it has Gaussian distribution with zero mean and covariance matrix $(\mathbf{S}^{-1} + \mathbf{R}^\dagger \mathbf{N}^{-1} \mathbf{R})^{-1}$. This property is the basis of the constrained realization method [6].

Another estimator can be formulated from the point of view of Bayesian statistics. The main objective of this approach is to calculate the *posterior* probability of the model given the data, which is written according to Bayes' theorem as $P(\text{model}|\text{data}) \propto P(\text{data}|\text{model}) P(\text{model})$. The estimator of the underlying field (*i.e.*, model, in Bayes' language) is taken to be the one that maximizes $P(\text{model}|\text{data})$, which is the most probable field. One can show that in the case in which the *prior* assumed to be a Gaussian, the Bayesian estimator coincides with the $\mathbf{s}^{MV}$ ([20], [8], [17]).

## 3 Spherical Harmonics Expansion and SVD

Spherical harmonics (SH) have been used to probe the large scale structure from wide angle galaxy surveys ([11],[12],[15],[16]). These analyses consist of expanding the angular galaxy distribution in a set of spherical harmonics which form an orthonormal functional basis when the expansion is carried out over the full, $4\pi$, sky. The expansion coefficients in SH basis constitute a statistically orthogonal set (like the case of k-space), which makes it more comfortable to deal with the data in the case of full sky coverage.

Consider an underlying angular density field, given in terms of its spherical harmonic expansion, $a_{lm} = \int d\hat{\mathbf{r}}\, \mathcal{S}(\hat{\mathbf{r}}) Y^*_{lm}(\hat{\mathbf{r}})$, where the projected surface density is given by $\mathcal{S}(\hat{\mathbf{r}}) = \sum_{lm} a_{lm} Y_{lm}(\hat{\mathbf{r}})$. This field is sampled by a finite discrete distribution of galaxies, which suffers basically from incomplete sky coverage and shot-noise. The observed harmonics $c_{lm}$ are related to the underlying whole sky harmonics $a_{lm}$ by: $c_{lm} = \sum_{l'm'} W_{ll'}^{mm'} \{a_{l'm'} + \sigma_{l'm'}\}$ where the monopole term ($l' = 0$) is excluded [15]. $\sigma_{lm}$ is the shot-noise in the 'true' number-weighted harmonics $a_{lm}$'s. The noise variance is estimated as $\langle \sigma^2 \rangle = \mathcal{N}$ (the mean number of galaxies per steradian, independent of $l$ in this case). Notice that $W_{ll'}^{mm'}$, the harmonic transform of the mask, introduce 'cross-talk' between different harmonics.

Applying equation 3 to this case, it can be shown that the solution for this inversion is ([9], [20])

$$\mathbf{a}^{\text{MV}} = diag\left\{\frac{A_l}{A_l + \mathcal{N}}\right\} \mathbf{W}^{-1} \mathbf{c} \equiv \mathbf{B}^{-1} \mathbf{c}, \qquad (4)$$

where the vectors $\mathbf{a}$ and $\mathbf{c}$ represent the sets $\{a_{lm}\}$ and $\{c_{lm}\}$. Here $A_l$ is the cosmic variance in the harmonics, which is determined by the assumed power spectrum (*i.e.*, the *prior*).

Here, this method is applied to a standard CDM simulation characterized by $\Omega_0 h = 0.5$ with particles selected to mimic the IRAS 1.2 Jy galaxy catalog with ZOA $|b| = 15°$ (see [9] for the case of the true IRAS 1.2 Jy data with ZOA $|b| = 5°$). The simulation evolves the particles until the rms variance of the density field in a sphere of $8 Mpc/h$ reache $\sigma_8 = 0.62$ (see [5] for more details). Figure 1 shows the harmonic reconstruction of the projected counts of the simulation using the raw harmonic coefficients, $c_{lm}$, up to $l_{max} = 15$. In this case the direct inversion of the matrix **B** is unstable and yields excessive power on small scales. However using the SVD algorithm to invert the matrix **B** sheds a new light on the question of the amount of useful information contained in the data.

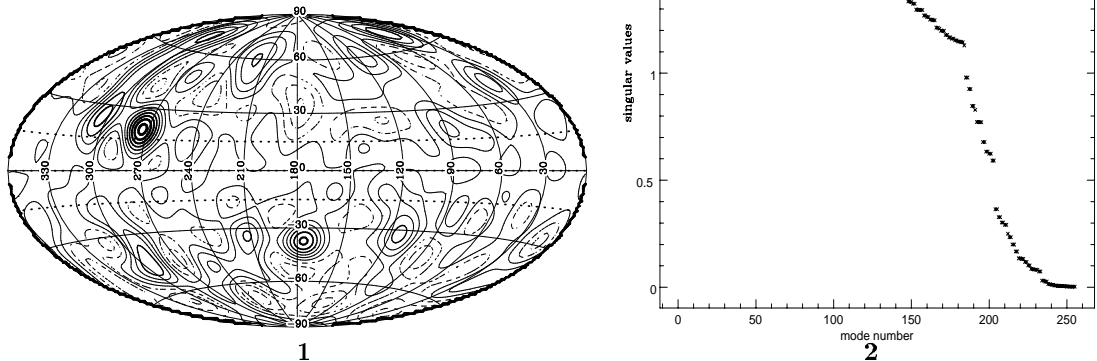

**Figure 1**: Harmonics reconstruction of the projected counts of the mock IRAS 1.2 Jy catalog with a 'Zone of Avoidance' $|b| = 15°$, in Galactic Aitoff projection, using the observed coefficients $c_{lm}$, up to $l_{max} = 15$. The contour levels of the projected surface number density are in steps of 100 galaxies per steradian (the mean projected density is $\mathcal{N} \sim 400$ galaxies per steradian).

**Figure 2**: The sorted spectrum of the singular values of the matrix **B**, for the $|b| = 15°$ case. The 'knee' suggests how to choose the cutoff for the SVD, $\lambda_{min}/\lambda_{max} = 0.565$, at the mode 184 ($l_{max} \approx 13$).

Essentially, the SVD algorithm decompose the square matrix **B** into the product of three matrices $\mathbf{B} = \mathbf{U}\, diag\{\lambda_i\}\, \mathbf{U^T}$ (SVD can also be applied to non-square matrices [13]). The $\lambda_i$'s and the rows of the matrix **U**, are the eigen-values and eigen-vectors of the matrix **B**. After performing the decomposition, the inversion of **B** is trivial,

$$\mathbf{B}^{-1} = \mathbf{U}^T\, diag\{1/\lambda_i\}\, \mathbf{U}. \tag{5}$$

Formally speaking, equation 4 has a unique solution if and only if **B** is a non-singular matrix, namely if $\lambda_i \neq 0$ for all $i$. However a meaningful solution to equation 4 can be obtained even in the case where **B** is singular, by requiring the solution to minimize the norm of the residuals, $|\mathbf{Ba} - \mathbf{c}|$. Such a solution is obtained by substituting $1/\lambda_i = 0$ in the expression for the inverse (equation 4) for any $\lambda_i = 0$ ([13]). Realistically, $1/\lambda_i$ is set to zero for any $\lambda_i$ less than some lower limit, determined by the physical and geometrical properties of the system. The question of how, in general, one should set the lower limit is discussed in §4 . In general, the eigenvalues measure the amount of 'information' carried by each mode in the problem; the small eigenvalues are those that might destabilize the inversion, so they do notcontribute significantly to the reconstruction.

Figure 2 shows the sorted spectrum of the eigenvalues ($\lambda_i$) of **B**, versus the harmonic number, $l$. The 'knee' in this figure suggests that the cutoff for the SVD, should be at $\lambda_{min} = 0.565\lambda_{max}$. Note that this cutoff suggest that only the largest 184 modes are significant ($l \approx 13$). Figure 3 shows the reconstructed $a_{lm}$'s map using WF and the above cutoff for SVD. Note that here, contrary to the $|b| = 5°$ case discussed by Lahav et al. [9] where the structure was recovered in the ZOA, the reconstructed ZOA remains empty. This illustrates that WF, even with the use of SVD extra-regularization, can not create structures out of nothing, unless the structures are dictated by the correlations ([9]).

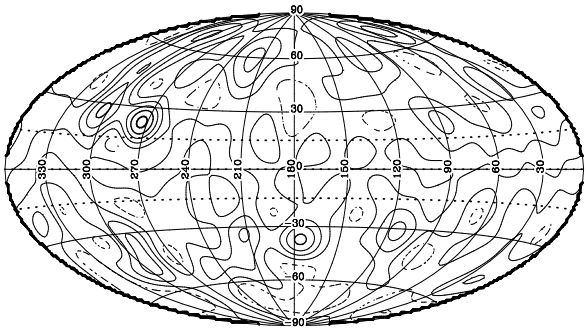 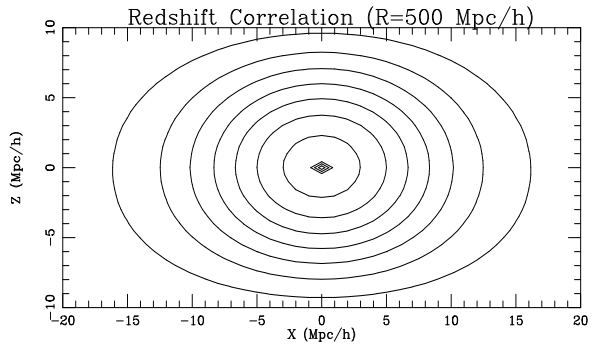

**Figure 3**: The reconstructed $a_{lm}$'s using Wiener filter and the cutoff shown in figure 2

**Figure 4**: The correlation function for SCDM model with $8Mpc/h$ smoothing, as measured in the redshift space according to the exact linear calculation around point at $500Mpc/h$ distance. Z is the *l.o.s.* direction.

## 4 'Optimal' Functional Basis

Frequently, it is useful to expand the cosmological perturbation field as a linear combination of a complete orthogonal functional basis. Such a basis defines a space in which all of the physical, statistical and geometrical properties of the field are retained. In the framework of linear gravitational instability, spaces of a special interest are Fourier space ([12]) and the space defined by SH and spherical Bessel functions ([4],[20]). These spaces are commonly used for the following reasons: a) Their expansion coefficients are statistically orthogonal (*e.g.*, $\langle \delta_{\mathbf{k}} \delta_{\mathbf{k}'} \rangle = P(k)\delta_D(\mathbf{k} - \mathbf{k}')$). b) They are eigen-functions of the Laplacian operator (and the $\nabla$ operator, in case of Fourier space).

However, the problems from which galaxy catalogs suffer complicate the picture since they introduce cross-talk between different modes, both in the signal covariance matrix itself and in the noise correlation matrix, as shown in §3 just above equation 4. For instance, in the linear regime, the redshift distortion, due to its anisotropy 'squeezes' the configuration space correlation function along the line of sight. Figure 4 shows the correlation matrix in redshift configuration space, around $R = 500Mpc/h$, for SCDM model in the framework of linear theory ([7],[19]). In k-space this is reflected in a cross-talk between different Fourier modes [19].

A different approach to the question of the functional basis, is to construct a statistically orthogonal functional basis according to our *prior* knowledge of the survey geometry, clustering properties and weighting scheme.

One possibility, which is useful for Wiener reconstruction, is to find the eigen-functions and eigen-values of the matrix $\mathbf{F}$ defined by equation 3. Indeed this is the case presented in §3, where the expansion was used to stabilize the inversion by means of data compression (*i.e.*, excluding small eigen-values). However, that was is a special case, in the sense that it had a characteristic scale (defined by the mask width) that corresponds to the position of the 'knee' in figure 2. Generally speaking, diagonalizing $\mathbf{F}$ does not guarantee such a 'natural' choice for the cutoff. Moreover, this expansion is not, necessarily, useful for other cases, especially for likelihood analysis.

Another possibility, is to find the eigen-functions of the matrix $\left\langle \mathbf{d}\,\mathbf{d}^\dagger \right\rangle$ (signal + noise)[10], which appears both in WF and in the Gaussian likelihood probability. However, the main problem of this approach, is that one can't distinguish between the contribution of the signal and the contribution of the noise to the eigen-values. Therefore it can not be used for data compression, nor for stabilizing the inversion, without throwing away useful information.

Here we suggest an 'improved SVD' approach, which basically solves the following generalized eigen-value problem,

$$\mathbf{R}\,\mathbf{S}\mathbf{R}^\dagger \mathbf{a}_i = \lambda_i \mathbf{N}\mathbf{a}_i \qquad (6)$$

where $\mathbf{a}_i$ and $\lambda_i$ are the eigen-vector and its eigen-value respectively. This equation diagonalizes $\mathbf{N}$ and $\mathbf{R\,S R}^\dagger$ simultaneously[1]. Note that the orthogonality condition of the vectors $\mathbf{a}$ is $\mathbf{N}$ weighted [18] (see also [2]).

This approach provides a natural way to compress the data, since here the eigen-values represent the relative contribution of the signal with respect to the noise in each mode. Hence, if $\lambda_i \geq 1$, the mode is dominated by the signal, and it is dominated by the noise otherwise. Note also that the eigen-vectors constitute a functional basis, although not complete, which includes all the *prior* knowledge about the data.

To summarize, this 'optimal' functional basis together with the WF formalism, the constrained realization method, and the Bayesian likelihood analysis (parameter estimation) provide a comprehensive and self contained prescription for analyzing the (linear) LSS of the Universe.

**Acknowledgements.** I am grateful to K. Fisher, Y. Hoffman, O. Lahav for their contribution to the work presented here and for many stimulating discussions

---

[1]This is similar to the problem of small oscillations in classical mechanics, where the noise and the signal matrices are analogous to the kinetic energy and potential energy matrices.